\documentclass[reprint,aps,prl,superscriptaddress,longbibliography,twocolumn,floatfix,10pt]{revtex4-2}

\usepackage{bm}
\usepackage{dcolumn}
\usepackage{amsthm}
\usepackage{amssymb}
\usepackage{amsmath}
\usepackage{graphicx}
\usepackage{bm}
\usepackage{xcolor}
\usepackage{fix-cm}
\usepackage{mathptmx} 
\usepackage[T1]{fontenc}
\usepackage[colorlinks,allcolors=blue]{hyperref}
\makeatletter
\def\NAT@def@citea{\def\@citea{\NAT@separator}}
\makeatother

\begin{document}

\title{Shell effects in quasi-fission for calcium induced reactions forming thorium isotopes}

\author{C. Simenel}\email{cedric.simenel@anu.edu.au}
\affiliation{Department of Fundamental and Theoretical Physics, and Department of Nuclear Physics and Accelerator Applications, Research School of Physics, The Australian National University, Canberra ACT  2601, Australia}
\author{A.S. Umar}\email{sait.a.umar@vanderbilt.edu}
\affiliation{Department of Physics and Astronomy, Vanderbilt University, Nashville, Tennessee 37235, USA}
\author{K. Godbey}\email{godbeyky@msu.edu}
\affiliation{Facility for Rare Isotope Beams, Michigan State University, East Lansing, Michigan 48824, USA}
\author{P. McGlynn}\email{mcglynn@frib.msu.edu}
\affiliation{Facility for Rare Isotope Beams, Michigan State University, East Lansing, Michigan 48824, USA}
\date{\today}


\begin{abstract}
Quantum shell effects induce an asymmetric fission mode in actinides, which disappears in neutron deficient isotopes.
Quasi-fission, characterized by a significant mass transfer in heavy ion collisions at low-energies, is expected to be affected by similar shell effects.
This is studied in $^{40-56}$Ca$+^{176}$Yb reactions with the time-dependent Hartree-Fock approach.
All reactions exhibit a mass equilibration process that stops when a heavy fragment with $Z\simeq54$ protons is formed.
Unlike the fission of thorium compound nuclei, quasi-fission does not exhibit a transition to symmetric modes in neutron deficient systems.
This observation is interpreted  in terms of potential energy surfaces that show a persistence of an asymmetric valley with an increasing barrier preventing its population in fission of the most neutron deficient thorium isotopes.
\end{abstract}
\maketitle

Quasi-fission occurring in heavy-ion collisions and fission of a compound nucleus are distinct processes.
In contrast to fission -- where an equilibrated compound nucleus decays into two fragments -- quasi-fission involves significant mass transfer between colliding nuclei without the formation of a fully equilibrated compound system~\citep{toke1985}. For a recent overview of quasi-fission experiments, see~\citep{hinde2021}.


Shell effects are expected to play a central role in fragment formation for both processes.
In fission, they are responsible for the preference toward asymmetric mass splits, particularly the  production of heavy fragments around $Z \simeq 54$ in actinides~\citep{schmidt2000,bockstiegel2008,chatillon2019}, a feature linked to octupole-deformed shell closures at $Z = 52$ and $56$~\citep{scamps2018}.
In quasi-fission, shell effects can inhibit mass equilibration between the fragments, thereby influencing the fragment mass distribution \cite{itkis2004,nishio2008,kozulin2010,nishio2012,kozulin2014,wakhle2014,itkis2015,morjean2017,hinde2018,kozulin2021,pal2024}.

Experimental mass distributions in quasi-fission reactions with an actinide target often show peaks near the doubly magic nucleus $^{208}$Pb, suggesting the stabilizing role of closed shells in this region~\citep{itkis2004,nishio2008,wakhle2014,morjean2017,hinde2018}. These experimental results are supported by microscopic calculations~\citep{wakhle2014,umar2016,mcglynn2023}. Recent experimental work provided {\it "clear evidence of quantum shell effects in slow quasi-fission processes,"} showing enhanced production of fragments near $A \approx 96$~\citep{pal2024}. This observation is consistent with theoretical predictions emphasizing the role of the $N = 56$ octupole-deformed shell gap~\citep{godbey2019}. However, alternative interpretations have been proposed where it is suggested that the observed $A \simeq 208$ peak might result from sequential fission of the target-like fragment, reducing the amount of binary events for masses between $A=208$ and the target mass~\citep{jeung2022}.
Reactions involving sub-lead targets offer an opportunity to disentangle shell effects from possible sequential fission.
However, the interpretation of shell effects in such reactions remains a topic of ongoing investigation.
Indeed, some experiments report {\it "clear evidence"} of shell-stabilized fragments~\citep{chizhov2003}, while others find only {\it "weak evidence"}~\citep{hinde2022}.

To further explore the parallels between fission and quasi-fission, recent studies compared static mean-field calculations for fission in $^{226}$Th with time-dependent mean-field simulations of quasi-fission, suggesting similar underlying shell effects in both processes~\citep{simenel2021,lee2024b}. However, this study was limited to a single compound nucleus, while the impact of shell effects in fission depends on the isotope, as illustrated by the transition from asymmetric to symmetric fission: Symmetric (resp. asymmetric) fission is dominant in thorium isotopes with $A<226$ ($A>226$) \cite{schmidt2000,bockstiegel2008,chatillon2019}. Whether or not a similar transition is present in quasi-fission is the question we address in this letter by performing microscopic simulations of $^A$Ca$+^{176}$Yb collisions for even mass numbers $40\le A\le 56$. The compound nuclei for these reactions are thorium isotopes with $216\le A\le232$, thus spanning the symmetric-to-asymmetric transition.

We use the time-dependent Hartree-Fock (TDHF) theory with a Skyrme energy density functional (EDF) describing both structure and dynamics of nuclear systems.
This approach has been applied to a broad range of systems to study quasi-fission \cite{wakhle2014,oberacker2014,ayik2015a,umar2015a,hammerton2015,sekizawa2016,umar2016,morjean2017,yu2017,guo2018c,zheng2018,godbey2019,li2019,simenel2021,li2022,stevenson2022,mcglynn2023,lee2024b,scamps2024,li2024c} (see \cite{simenel2025} for a recent review).
Notably, the approach involves no free parameters, relying solely on the Skyrme EDF as its phenomenological input. The parameters of the EDF are typically constrained by properties of selected nuclei and infinite nuclear matter. In this work, we use the SLy4$d$ parametrization \cite{kim1997}, suitable for both static calculations and simulations of heavy-ion collisions.

The \textsc{tdhf3d} code is used with a plane of symmetry (the reaction plane) \cite{kim1997}.
BCS pairing correlations are included in the ground-states with density dependent delta interaction and the occupation numbers are kept constant during the time evolution according to the frozen occupation approximation.
The  $^{176}$Yb ground-state is prolate with  deformation parameter $\beta_2\approx0.33$, while the calcium isotopes are spherical.
The center of mass energy of the reaction $E_\mathrm{c.m.}=172$~MeV corresponds to $10-15\%$ above the Coulomb barrier \cite{swiatecki2005} (see Table 1 in supplemental material \cite{supplemental}).
The initial distance between the centers of mass of the nuclei is $22.6$~fm.
Four initial orientations of the $^{176}$Yb (whose deformation axis is in the reaction plane, initially at an angle of 0, 45, 90, and 135 degrees with respect to the axis joining the initial centers of mass) are considered to account for the effect of orientation on reaction mechanism.
A Cartesian grid of $72\times72\times(28/2)\times\Delta x^3$ is used with a mesh size $\Delta x=0.8$~fm.
The calculations are stopped when two separate fragments are formed and their centers of mass are separated by 26~fm.
If the compound system is still present after an evolution of at least 30~zs, and its elongation is not increasing, then fusion is assumed.
For each orientation, quasi-fission occurs in a range of orbital angular momenta $L$ below which the nuclei fuse and above which (in)elastic scattering occurs.
The step in $L$ is set to $\Delta L=2\hbar$ in this range.
The contact time $\tau$ is defined as  the time during which the density exceeds 0.08~fm$^{-3}$ (approximately half the nuclear saturation density) in the neck between the fragments, or when the neck cannot be defined due to the compact shape of the compound system.
The number of protons ($Z$) and neutrons ($N$) in the primary fragments (i.e., prior to subsequent decay) are evaluated from integrating the proton and neutron densities, respectively, in the last TDHF iteration.
The results of  480 TDHF calculations are compiled in supplemental material \cite{supplemental}.

\begin{figure}[!htb]
\centerline{\includegraphics[width=8cm]{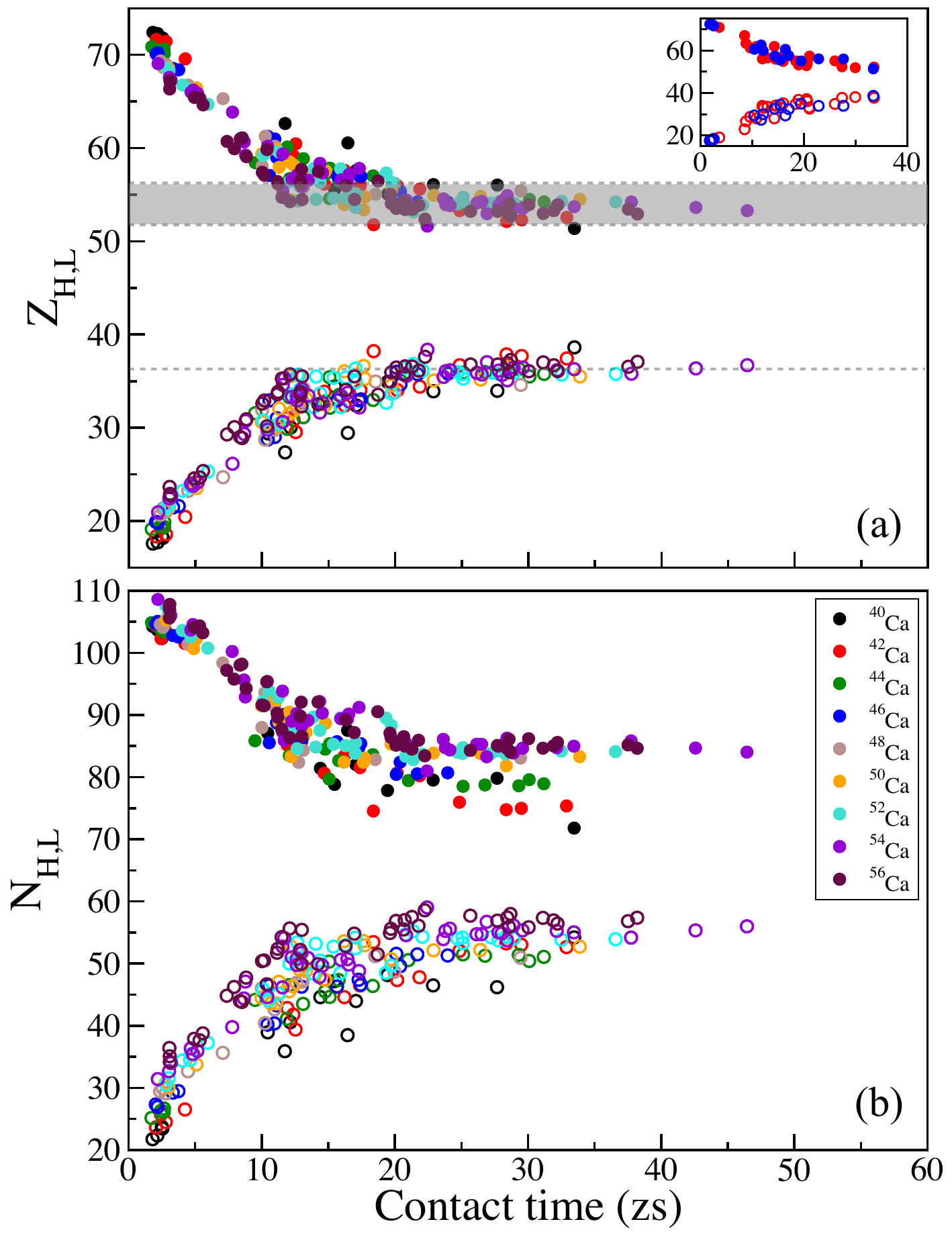}}
\caption{Number of (a) protons ($Z_{H,L}$) and (b) neutrons ($N_{H,L}$) in the fragments formed in $^{40-56}$Ca$+^{176}$Yb quasi-fission reactions  at $E_{c.m.}=172$~MeV as function of contact times given in zeptoseconds (1~zs$=10^{-21}$~s).
 Full and open circles correspond to heavy ($H$) and light ($L$) fragments, respectively. The color code for the calcium isotopes shown in (b) is the same for both figures.
 The inset in (a) shows results for $^{40}$Ca$+^{176}$Yb at $E_{c.m.}=172$~MeV (blue symbols) and $E_{c.m.}=180$~MeV (red symbols). The octupole deformed proton shell effects at $Z=52-56$ are shown with a gray band. The deformed $Z=36$ proton shell is shown with dashed line. }
\label{fig:NZ_T}
\end{figure}

Figure \ref{fig:NZ_T} shows the number of protons and neutrons in the fragments formed in quasi-fission reactions as a function of contact times.
Remarkably, all reactions exhibit a similar behavior, with a partial mass-equilibration that stops after $15-20$~zs contact time.
This relatively slow mass equilibration process is characteristic to quasi-fission \cite{toke1985,durietz2013,simenel2020}.
Importantly, mass transfer is stopped when a heavy fragment with $Z_H\simeq54$ protons is formed, as in the asymmetric mode in actinide fission.
This corresponds to a light fragment with $Z_L\simeq36$ which has been identified as a stabilizing deformed shell in light fragments  in sub-lead fission \cite{morfouace2025,buete2025}.
However, final neutron numbers are more spread, indicating that they are less likely to drive the final asymmetry.
The inset of Fig.~\ref{fig:NZ_T} also shows that similar behaviors are found at two different energies corresponding to 10\% and 15\% above the Coulomb barrier in the $^{40}$Ca$+^{176}$Yb reaction.
Thus, our conclusions would not be affected by a small increase in energy.

\begin{figure}[!htb]
\centerline{\includegraphics[width=8cm]{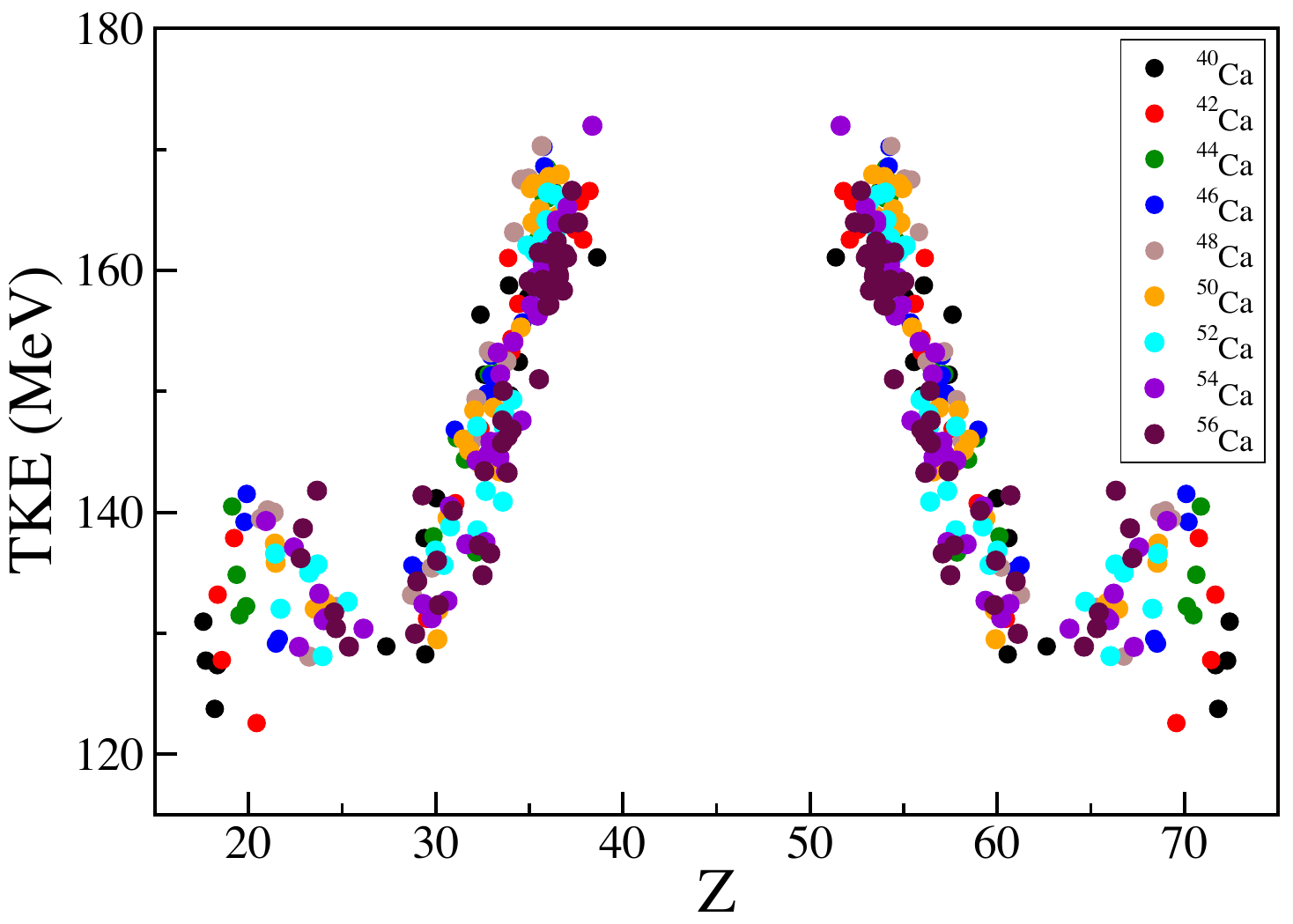}}
\caption{Total kinetic energy of the outgoing fragments of $^{40-56}$Ca$+^{176}$Yb quasi-fission reactions at $E_{c.m.}=172$~MeV as a function of the number of protons $Z$ in the fragments. }
\label{fig:TKE}
\end{figure}

The total kinetic energy (TKE) of the outgoing fragments provides a proxy for the shape of the system at scission.
Indeed, a compact shape leads to a larger TKE than an elongated configuration due to stronger Coulomb repulsion between nascent fragments.
The TKE can be computed from the sum of the kinetic energy of the fragments and their Coulomb potential energy (assuming point like fragments) after scission \cite{simenel2014a,scamps2015a}.
Figure~\ref{fig:TKE} shows the resulting TKE as a function of the number of protons in the fragments.
The strong correlation observed for $30\lesssim Z\lesssim60$ is interpreted as an increase of Coulomb repulsion between more symmetric fragments.
The overall strong similarities between the systems demonstrate that their shapes in the exit channel are similar for all reactions and a given final charge asymmetry, despite the entrance channel being different in terms of neutron numbers and orientations.

\begin{figure*}[!htb]
\centerline{\includegraphics[width=16cm]{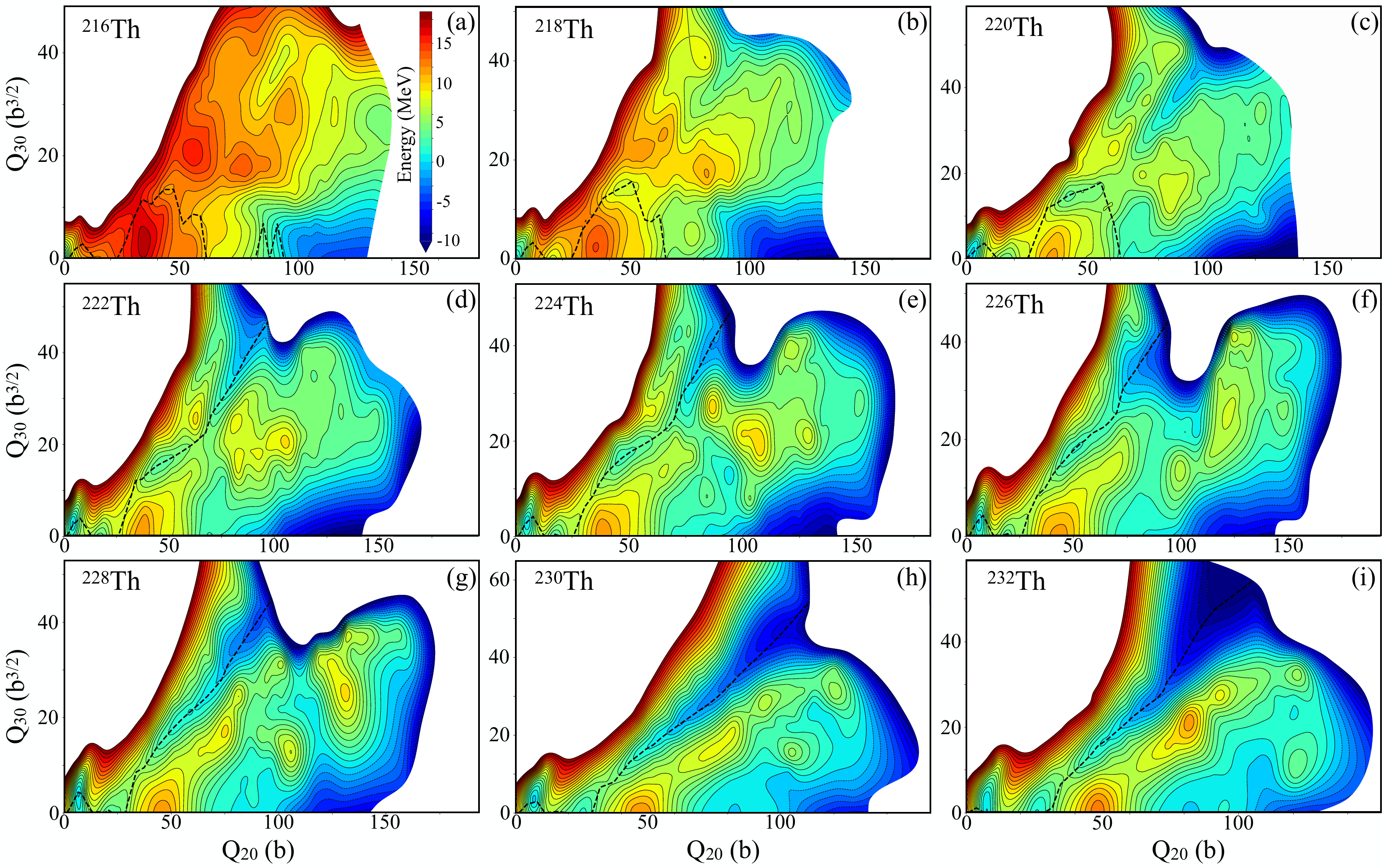}}
\caption{Potential energy surfaces of $^{216-232}$Th even $A$ isotopes. The energy scale shown in (a) is the same for all PES.
The dashed lines show the fission path obtained by leaving the octupole moment free.}
\label{fig:PES-Th}
\end{figure*}

The fact that all studied reactions produce similar asymmetric fragments for contact times exceeding $15-20$~zs implies that the asymmetric-to-symmetric transition experimentally observed in fission of thorium isotopes is not predicted to be a feature of quasi-fission.
In fission, this transition occurs around $^{226}$Th \cite{schmidt2000}.
To interpret this difference, potential energy surfaces (PES) of even isotopes of thorium have been computed for $216\le A\le232$ with the constrained Hartree-Fock  method with BCS pairing.
The \textsc{skyax} code \cite{reinhard2021} was used with the same EDF as the TDHF calculations.
Constraints are imposed on elongation through the quadrupole moment
\begin{equation}
    Q_{20} = \sqrt{\frac{5}{16\pi}}\int d^{3}{r}\; \rho(\textbf{r})(2z^2-x^2-y^2),
\end{equation}
and on asymmetry with the octupole moment
\begin{equation}
    Q_{30} = \sqrt{\frac{7}{16\pi}}\int d^{3}{r}\; \rho(\textbf{r})(2z^3-3z(x^2+y^2)).
\end{equation}
The \textsc{skyax} code assumes axial symmetry. The states are also built with no internal excitation and zero average angular momentum.
Although these assumptions are usually broken in quasi-fission reactions, PES have been successfully used to interpret quasi-fission trajectories in the $Q_{20}-Q_{30}$ plane \cite{lee2024b,mcglynn2023}.
\begin{figure*}[!htb]
    \centerline{\includegraphics[width=18cm]{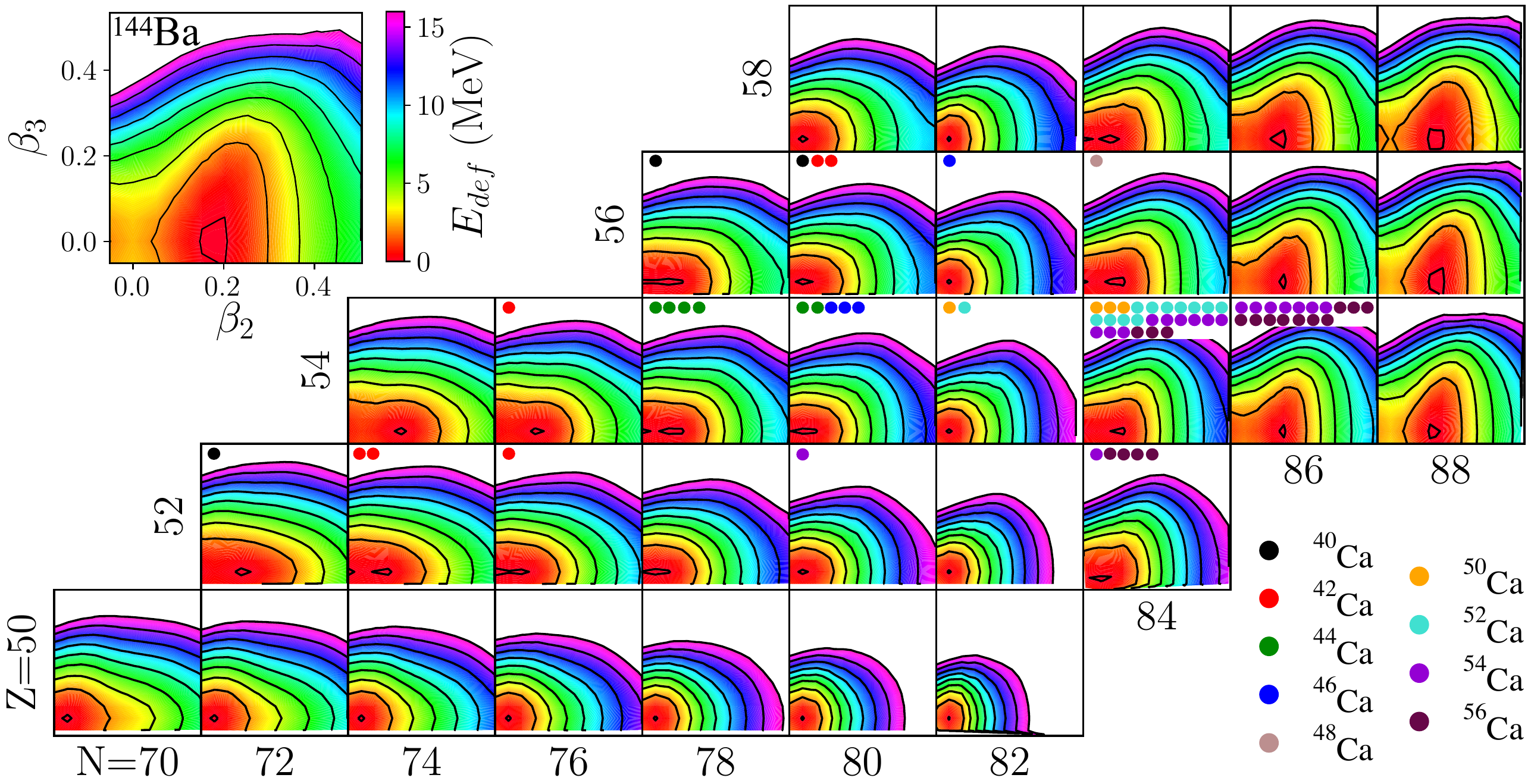}}
    \caption{Potential energy surfaces in the $\beta_2-\beta_3$ plane for even $Z$ and even  $N$ nuclei. The axis legends and color code for the deformation energy $E_{def}$ are the same for each PES and detailed in the case of $^{144}$Ba.
        The circles represent the closest even-even heavy quasi-fission fragment in $^{40-56}$Ca$+^{176}$Yb for contact times $\tau>20$~zs.}
    \label{fig:frag}
\end{figure*}

The PES for thorium isotopes  shown in Fig.~\ref{fig:PES-Th}(a-i) have been obtained  with quadrupole and octupole steps $\Delta Q_{20}=1.44$~b and $\Delta Q_{30}=1$~b$^{3/2}$, respectively.
The minimum energy fission path is obtained by only constraining $Q_{20}$ and increasing its value from the ground-state to scission.
Scission is defined by configurations where the neck density drops below $\rho=0.08$~fm$^{-3}$.

The $^{232}$Th PES [Fig.~\ref{fig:PES-Th}(i)] is typical of what is expected for an actinide near the valley of stability.
Starting from the ground-state, the minimum energy fission path overcomes a first barrier along the symmetric path ($Q_{30}=0$) to reach a second minimum.
It then acquires an asymmetry due to neutron shell effects in the compound system \cite{gustafsson1971,brack1972,bernard2023}.
After overcoming a fission saddle point, the system falls into the asymmetric fission valley towards scission.
In particular, a barrier ridge prevents the fissioning nucleus from returning to the symmetric fission valley.

Moving towards lighter thorium isotopes, the asymmetric valley is pushed towards larger elongations and asymmetries.
In $^{216}$Th, the most neutron deficient isotope, the asymmetric valley only remains as a shallow pocket in the PES.
Although the minimum energy fission path remains in the asymmetric valley down to $^{222}$Th, it is clear that the system may return to the symmetric valley without passing a third barrier already in $^{226}$Th \cite{simenel2021}.
From $^{220}$Th downward in mass, this path effectively returns to symmetry after going around the second barrier.

These observations are in qualitative agreement with experiment \cite{schmidt2000,bockstiegel2008,chatillon2019}.
The persistence of the asymmetric valley across these thorium isotopes is also interpreted as the reason for the observed asymmetric quasi-fission in TDHF calculations.
Indeed, in $^{40-56}$Ca$+^{176}$Yb quasi-fission, the entrance channel is more asymmetric than the outgoing fragments.
A possible explanation for asymmetric quasi-fission is that, after the initial kinetic energy is dissipated into internal excitations, the systems fall into the asymmetric valley, leading to the production of heavy fragments with $Z\simeq54$ protons.

Figure \ref{fig:frag} shows the PES computed with \textsc{skyax} in the region of the heavy quasi-fission fragments produced in $^{40-56}$Ca$+^{176}$Yb. The deformation energy $E_{def}$ is plotted as a function of the dimensionless quadrupole and octupole deformation parameters $\beta_2$ and $\beta_3$, respectively.
Except for magic nuclei ($Z=50$ or $N=82$), the nuclei are able to acquire an elongation ($\beta_2>0$) and, for $N\ge84$, an asymmetry ($\beta_3\ne0$) for no or little cost in deformation energy.
As in the case of fission, the production of such fragments in quasi-fission is expected to be favored.
Indeed, such deformations are induced at scission by the competition between Coulomb repulsion and strong attraction in the neck \cite{scamps2018}.

Figure \ref{fig:frag} also shows the nearest even-even heavy fragments produced for each reaction with $\tau>20$~zs.
We indeed observe that the magic number $Z=50$ and, to a lesser extent, $N=82$, are avoided, which is compatible with the interpretation that the large deformation energy in these magic nuclei hinders the production of such fragments.
We also note that the more neutron rich systems have a greater chance to reach $\tau>20$~zs.
A possible interpretation is that such systems have a longer asymmetric fission valley (see Fig.~\ref{fig:PES-Th}) and the quasi-fissioning system takes more time to reach scission.

In this work, we have studied the role of shell effects in quasi-fission with TDHF by varying the neutron number of the reactants.
Unlike the compound nuclei (thorium isotopes) that undergo an asymmetric-to-symmetric fission transition by removing neutrons, quasi-fission is found to populate the asymmetric mode with $Z\simeq54$ protons in all systems.
This difference between fission and quasi-fission is interpreted in terms of the PES of the compound system, as well as that of the heavy fragments.
We conclude that the asymmetric valley is still present in neutron deficient thorium isotopes, and is responsible for the asymmetric quasi-fission mode, while large barriers prevent  such systems from reaching the asymmetric valleys in fission.

\begin{acknowledgments}
This work has been supported by the Australian Research Council Discovery Project (project number DP190100256) funding schemes, by the U.S. Department of Energy under grant Nos. DE-SC0013847 (Vanderbilt University) and DE-SC0013365 (Michigan State University), and by the NNSA Cooperative Agreement DE-NA0004245.
This work was also supported by computational resources provided by the Australian Government through the National Computational Infrastructure (NCI) under the ANU Merit Allocation Scheme.
\end{acknowledgments}
\bibliography{VU_bibtex_master.bib}

\newpage

\section*{APPENDIX: supplemental material}

The results presented in this supplemental material are for TDHF calculations of $^{40,42,44,46,48,50,52,54,56}$Ca$+^{176}$Yb collisions at $E_{c.m.}=172$~MeV and for $^{40}$Ca$+^{176}$Yb at $E_{c.m.}=180$~MeV.
Table~\ref{tab:barriers} lists the Coulomb capture barriers $V_B$ for  these systems as estimated from the systematic of Swiatecki et al. \cite{swiatecki2005}. 

\begin{table}[ht]
 \caption{\label{tab:barriers} Coulomb capture barriers \cite{swiatecki2005} in $^{A}$Ca$+^{176}$Yb reactions.}
 \begin{ruledtabular}
 \begin{tabular}{lccc}
$A$ & $V_B$ (MeV) & $E_{c.m.}$ (MeV)& $E_{c.m.}/V_B$ \\
\colrule
\colrule
40 & 156.7 &  180 & 1.15 \\
\colrule
40 &  156.7&  172 & 1.10  \\
42 &  155.6&   & 1.11 \\
44 &  154.6& &  1.11\\
46 &  153.6& & 1.12\\
48 &  152.7& &  1.13\\
50 &  151.8& & 1.13 \\
52 &  150.9& & 1.14 \\
54 &  150.1& &  1.15\\
56 &  149.3& &1.15  
  \end{tabular}
 \end{ruledtabular}
 \end{table}

\noindent
In each table, the results are grouped according to initial orientation: the $^{176}$Yb deformation axis has an angle of 0, 45, 90 or 135 degrees with the line connecting the centre of masses of the nuclei in the TDHF initial condition. 
The angle $\beta$ is the angle (in degrees) between the deformation axis of $^{176}$Yb and its velocity vector at the initial TDHF condition. 
The initial orbital angular momentum  $L$ is given in units of $\hbar$.
The contact time $\tau$ is given in zeptoseconds (1~zs$=10^{-21}$~s). 
Only results with $\tau\gtrsim2$~zs are reported in the tables.
The subscript $H$ and $L$ stand for heavy and light fragment, respectively. 
The total kinetic energy (TKE)  is given in MeV.


  \begin{table}[ht]
 \caption{$^{40}\mbox{Ca}+^{176}\mbox{Yb}$ at $E_{c.m.}=180$~MeV. \label{tab:40Ca180MeV}}
 \begin{ruledtabular}
 \begin{tabular}{lccccccc}
$\beta$ (deg.) & $L$ ($\hbar$) & $\tau$ (zs)& $Z_H$ & $N_H$ & $Z_L$ & $N_L$ & TKE (MeV)\\
\colrule
       &  60-66 &  >35.91 &        &        &        &        &        \\
 14.9 &  68 &   29.97 &  51.93 &  72.84 &  38.07 &  53.16 & 163.47 \\
 15.3 &  70 &   33.60 &  52.21 &  73.52 &  37.79 &  52.48 & 162.20 \\
 15.8 &  72 &   18.42 &  55.16 &  78.65 &  34.84 &  47.35 & 163.26 \\
 16.2 &  74 &   12.02 &  56.15 &  79.82 &  33.85 &  46.18 & 158.74 \\
 16.7 &  76 &   14.53 &  56.03 &  79.94 &  33.97 &  46.06 & 152.32 \\
 17.1 &  78 &   11.98 &  56.10 &  79.36 &  33.90 &  46.64 & 157.54 \\
 17.6 &  80 &   12.99 &  56.54 &  80.14 &  33.46 &  45.86 & 154.98 \\
 18.0 &  82 &   10.81 &  61.92 &  88.75 &  28.08 &  37.25 & 136.29 \\
 18.5 &  84 &    9.63 &  61.27 &  88.28 &  28.73 &  37.72 & 134.27 \\
 18.9 &  86 &    2.60 &  71.64 & 101.82 &  18.36 &  24.18 & 126.41 \\
\colrule
  &  72-76 &  >31.31 &        &        &        &        &        \\
 61.9 &  78 &   27.41 &  52.30 &  73.73 &  37.70 &  52.27 & 162.16 \\
 62.3 &  80 &   20.53 &  52.87 &  74.99 &  37.13 &  51.01 & 164.93 \\
 62.8 &  82 &   20.67 &  53.88 &  75.50 &  36.12 &  50.50 & 163.20 \\
 63.2 &  84 &   21.03 &  56.81 &  80.64 &  33.19 &  45.36 & 164.92 \\
 63.7 &  86 &   16.01 &  54.87 &  77.91 &  35.13 &  48.09 & 155.73 \\
 64.1 &  88 &   15.58 &  56.92 &  80.80 &  33.08 &  45.20 & 156.63 \\
 64.5 &  90 &    8.80 &  63.35 &  90.76 &  26.65 &  35.24 & 133.47 \\
 65.0 &  92 &    3.65 &  70.95 & 101.43 &  19.05 &  24.57 & 125.39 \\
\colrule
  &  44-48 &  >31.04 &        &        &        &        &        \\
101.4 &  52 &   19.00 &  53.23 &  75.42 &  36.77 &  50.58 & 157.58 \\
101.8 &  54 &  >31.04 &        &        &        &        &        \\
102.2 &  56 &   26.03 &  55.16 &  78.45 &  34.84 &  47.55 & 151.42 \\
102.7 &  58 &  >31.04 &        &        &        &        &        \\
103.1 &  60 &   21.12 &  57.39 &  81.55 &  32.61 &  44.45 & 159.52 \\
103.6 &  62 &  >31.04 &        &        &        &        &        \\
104.0 &  64 &   12.28 &  57.00 &  80.82 &  33.00 &  45.18 & 155.40 \\
104.5 &  66 &   14.65 &  55.88 &  79.44 &  34.12 &  46.56 & 152.96 \\
104.9 &  68 &   14.31 &  61.94 &  89.07 &  28.06 &  36.93 & 131.25 \\
105.3 &  70 &    8.55 &  67.06 &  95.69 &  22.94 &  30.31 & 120.17 \\
105.8 &  72 &    2.36 &  72.11 & 102.58 &  17.89 &  23.42 & 124.88 \\
\colrule
 &  54-64 &  >31.23 &        &        &        &        &        \\
149.2 &  66 &    2.19 &  72.49 & 103.83 &  17.51 &  22.17 & 130.84 \\
 \end{tabular}
 \end{ruledtabular}
 \end{table}

  \begin{table}[ht]
 \caption{$^{40}\mbox{Ca}+^{176}\mbox{Yb}$ at $E_{c.m.}=172$~MeV. \label{tab:40Ca}}
 \begin{ruledtabular}
 \begin{tabular}{lccccccc}
$\beta$ (deg.) & $L$ ($\hbar$) & $\tau$ (zs)& $Z_H$ & $N_H$ & $Z_L$ & $N_L$ & TKE (MeV)\\
\colrule
  &  60-62 &  >31.40 &        &        &        &        &        \\
 14.7 &  64 &   19.45 &  55.07 &  77.85 &  34.93 &  48.15 & 157.77 \\
 15.2 &  66 &  >31.40 &        &        &        &        &        \\
  15.6 &  68 &   15.47 &  55.56 &  78.82 &  34.44 &  47.18 & 152.43 \\
  16.1 &  70 &   14.41 &  57.41 &  81.40 &  32.59 &  44.60 & 151.38 \\
  16.6 &  72 &   12.15 &  59.97 &  85.44 &  30.03 &  40.56 & 141.17 \\
  17.0 &  74 &   11.74 &  62.64 &  90.11 &  27.36 &  35.89 & 128.91 \\
  17.5 &  76 &    2.51 &  71.67 & 102.27 &  18.33 &  23.73 & 127.36 \\
\colrule
 &  60-70 &  >35.79 &        &        &        &        &        \\
 61.3 &  72 &   22.88 &  56.08 &  79.52 &  33.92 &  46.48 & 158.76 \\
 61.8 &  74 &   33.46 &  51.38 &  71.79 &  38.62 &  54.21 & 161.08 \\
 62.2 &  76 &  >35.79 &        &        &        &        &        \\
 62.7 &  78 &   17.08 &  57.61 &  82.02 &  32.39 &  43.98 & 156.34 \\
 63.2 &  80 &   10.46 &  60.60 &  87.06 &  29.40 &  38.94 & 137.87 \\
 63.6 &  82 &    2.17 &  72.28 & 103.67 &  17.72 &  22.33 & 127.74 \\
\colrule
 &  48-50 &  >31.01 &        &        &        &        &        \\
101.9 &  52 &   27.68 &  56.03 &  79.81 &  33.97 &  46.19 & 149.66 \\
102.4 &  54 &   16.45 &  60.56 &  87.53 &  29.44 &  38.47 & 128.25 \\
102.8 &  56 &    2.55 &  71.81 & 102.53 &  18.19 &  23.47 & 123.75 \\  
\colrule
 &  50-54 &  >31.21 &        &        &        &        &        \\
147.6 &  56 &    1.81 &  72.42 & 104.27 &  17.58 &  21.73 & 130.96 \\
 \end{tabular}
 \end{ruledtabular}
 \end{table}

  \begin{table}[ht]
 \caption{$^{42}\mbox{Ca}+^{176}\mbox{Yb}$ at $E_{c.m.}=172$~MeV. \label{tab:42Ca}}
 \begin{ruledtabular}
 \begin{tabular}{lccccccc}
$\beta$ (deg.) & $L$ ($\hbar$) & $\tau$ (zs)& $Z_H$ & $N_H$ & $Z_L$ & $N_L$ & TKE (MeV)\\
\colrule
  &  66-70 &  >31.40 &        &        &        &        &        \\
 16.2 &  72 &   20.18 &  55.96 &  80.67 &  34.04 &  47.33 & 153.28 \\
 16.7 &  74 &   16.21 &  57.61 &  83.40 &  32.39 &  44.60 & 146.91 \\
 17.2 &  76 &   11.92 &  58.95 &  85.16 &  31.05 &  42.84 & 140.76 \\
 17.6 &  78 &    4.24 &  69.57 & 101.48 &  20.43 &  26.52 & 122.58 \\
\colrule
  &  60-66 &  >31.28 &        &        &        &        &        \\
 60.1 &  68 &   28.36 &  52.12 &  74.76 &  37.88 &  53.24 & 162.55 \\
 60.5 &  70 &   29.50 &  52.30 &  74.97 &  37.70 &  53.03 & 165.69 \\
 61.0 &  72 &  >35.79 &        &        &        &        &        \\
 61.4 &  74 &  >35.79 &        &        &        &        &        \\
 61.9 &  76 &   32.89 &  52.56 &  75.36 &  37.44 &  52.64 & 163.37 \\
 62.3 &  78 &   14.70 &  56.13 &  80.63 &  33.87 &  47.37 & 161.03 \\
 62.8 &  80 &   24.85 &  53.30 &  75.95 &  36.70 &  52.05 & 163.80 \\
 63.2 &  82 &   18.39 &  51.78 &  74.54 &  38.22 &  53.46 & 166.56 \\
 63.7 &  84 &   12.54 &  60.46 &  88.62 &  29.54 &  39.38 & 131.19 \\
 64.2 &  86 &    2.07 &  71.65 & 104.37 &  18.35 &  23.63 & 133.19 \\
\colrule
 &  46-52 &  >31.00 &        &        &        &        &        \\
102.1 &  54 &   21.85 &  55.59 &  80.20 &  34.41 &  47.80 & 157.23 \\
102.6 &  56 &   17.39 &  55.95 &  81.52 &  34.05 &  46.48 & 154.33 \\
 103.0 &  58 &   12.38 &  59.16 &  86.23 &  30.84 &  41.77 & 140.30 \\
 103.5 &  60 &    2.44 &  70.77 & 102.27 &  19.23 &  25.73 & 137.87 \\
\colrule
 &  54-56 &  >31.19 &        &        &        &        &        \\
147.8 &  58 &    2.80 &  71.43 & 103.50 &  18.57 &  24.50 & 127.78 \\
  \end{tabular}
 \end{ruledtabular}
 \end{table}

 \begin{table}[ht]
 \caption{$^{44}\mbox{Ca}+^{176}\mbox{Yb}$ at $E_{c.m.}=172$~MeV.\label{tab:44Ca}}
 \begin{ruledtabular}
 \begin{tabular}{lccccccc}
$\beta$ (deg.) & $L$ ($\hbar$) & $\tau$ (zs)& $Z_H$ & $N_H$ & $Z_L$ & $N_L$ & TKE (MeV)\\
\colrule
  &  70-72 &  >31.40 &        &        &        &        &        \\
 16.4 &  74 &   18.35 &  57.02 &  83.61 &  32.98 &  46.39 & 151.35 \\
 16.8 &  76 &   13.11 &  58.87 &  86.50 &  31.13 &  43.50 & 146.13 \\
 17.3 &  78 &    9.50 &  58.44 &  85.85 &  31.56 &  44.15 & 144.37 \\
 17.7 &  80 &   11.90 &  60.12 &  88.90 &  29.88 &  41.10 & 138.01 \\
 18.2 &  82 &	 2.67 &  70.13 & 103.33 &  19.87 &  26.67 & 132.23 \\
\colrule
  &  60-70 &  >31.26 &        &        &        &        &        \\
 60.7 &  72 &   30.09 &  54.50 &  79.58 &  35.50 &  50.42 & 167.16 \\
 61.1 &  74 &   26.77 &  54.02 &  78.74 &  35.98 &  51.26 & 165.95 \\
 61.6 &  76 &   31.19 &  54.22 &  78.93 &  35.78 &  51.07 & 165.99 \\
 62.0 &  78 &  >33.52 &        &        &        &        &        \\
 62.5 &  80 &   21.02 &  54.06 &  79.45 &  35.94 &  50.55 & 168.49 \\
 62.9 &  82 &   15.07 &  54.58 &  79.68 &  35.42 &  50.32 & 162.47 \\
 63.3 &  84 &   29.31 &  53.71 &  78.60 &  36.29 &  51.40 & 166.56 \\
 63.8 &  86 &   12.08 &  56.73 &  83.46 &  33.27 &  46.54 & 150.20 \\
 64.2 &  88 &    2.65 &  70.64 & 103.86 &  19.36 &  26.14 & 134.83 \\
\colrule
 &  46-56 &  >30.99 &        &        &        &        &        \\
102.8 &  58 &   14.77 &  57.17 &  84.51 &  32.83 &  45.49 & 151.43 \\
103.2 &  60 &   15.78 &  56.48 &  82.67 &  33.52 &  47.33 & 151.99 \\
103.7 &  62 &   15.11 &  57.85 &  85.36 &  32.15 &  44.64 & 136.67 \\
104.1 &  64 &    2.43 &  70.48 & 103.74 &  19.52 &  26.26 & 131.49 \\
\colrule
147.6 &  58 &  >31.17 &        &        &        &        &        \\
148.0 &  60 &   25.12 &  54.09 &  78.53 &  35.91 &  51.47 & 157.10 \\
148.5 &  62 &    1.73 &  70.88 & 104.86 &  19.12 &  25.14 & 140.48 \\
 \end{tabular}
 \end{ruledtabular}
 \end{table}

 \begin{table}[ht]
 \caption{$^{46}\mbox{Ca}+^{176}\mbox{Yb}$ at $E_{c.m.}=172$~MeV. \label{tab:46Ca}}
 \begin{ruledtabular}
 \begin{tabular}{lccccccc}
$\beta$ (deg.) & $L$ ($\hbar$) & $\tau$ (zs)& $Z_H$ & $N_H$ & $Z_L$ & $N_L$ & TKE (MeV)\\
\colrule
  &  60-74 &  >31.40 &        &        &        &        &        \\
 16.5 &  76 &   20.40 &  55.36 &  82.41 &  34.64 &  49.59 & 155.70 \\
 17.0 &  78 &   11.11 &  58.98 &  88.72 &  31.02 &  43.28 & 146.82 \\
 17.4 &  80 &   10.96 &  60.99 &  91.64 &  29.01 &  40.36 & 135.22 \\
 17.9 &  82 &   10.44 &  61.25 &  91.83 &  28.75 &  40.17 & 135.62 \\
 18.3 &  84 &	 3.75 &  68.39 & 102.54 &  21.61 &  29.46 & 129.54 \\
 \colrule
  &  74-80 &  >31.26 &        &        &        &        &        \\
 62.6 &  82 &   21.69 &  54.24 &  80.53 &  35.76 &  51.47 & 170.21 \\
 63.0 &  84 &   20.11 &  53.99 &  80.44 &  36.01 &  51.56 & 167.26 \\
 63.5 &  86 &   23.97 &  54.20 &  80.70 &  35.80 &  51.30 & 168.62 \\
 63.9 &  88 &   15.66 &  57.04 &  85.70 &  32.96 &  46.30 & 151.31 \\
 64.3 &  90 &   13.00 &  57.25 &  85.72 &  32.75 &  46.28 & 149.83 \\
 64.8 &  92 &    2.19 &  70.22 & 105.06 &  19.78 &  26.94 & 139.21 \\
\colrule
  &  40-58 &  >30.98 &        &        &        &        &        \\
103.0 &  60 &   17.44 &  56.96 &  85.38 &  33.04 &  46.62 & 151.08 \\
103.4 &  62 &   17.28 &  57.05 &  84.62 &  32.95 &  47.38 & 152.94 \\
103.9 &  64 &   10.57 &  57.07 &  85.52 &  32.93 &  46.48 & 145.57 \\
 104.3 &  66 &	 3.33 &  68.54 & 102.77 &  21.46 &  29.23 & 129.15 \\
\colrule
 &  60-62 &  >31.17 &        &        &        &        &        \\
148.7 &  64 &    2.05 &  70.10 & 104.67 &  19.90 &  27.33 & 141.52 \\
  \end{tabular}
 \end{ruledtabular}
 \end{table}

 \begin{table}[ht]
 \caption{$^{48}\mbox{Ca}+^{176}\mbox{Yb}$ at $E_{c.m.}=172$~MeV. \label{tab:48Ca}}
 \begin{ruledtabular}
 \begin{tabular}{lccccccc}
$\beta$ (deg.) & $L$ ($\hbar$) & $\tau$ (zs)& $Z_H$ & $N_H$ & $Z_L$ & $N_L$ & TKE (MeV)\\
\colrule
 &  50-78 &  >31.39 &        &        &        &        &        \\
 17.1 &  80 &   13.00 &  57.07 &  86.71 &  32.93 &  47.29 & 153.23 \\
 17.6 &  82 &   10.25 &  61.27 &  93.59 &  28.73 &  40.41 & 133.17 \\
 18.0 &  84 &   10.97 &  60.21 &  91.37 &  29.79 &  42.63 & 135.43 \\
 18.4 &  86 &    7.08 &  65.31 &  98.37 &  24.69 &  35.63 & 132.18 \\
 18.9 &  88 &    2.58 &  68.66 & 104.09 &  21.34 &  29.91 & 139.98 \\
\colrule
 &  76-82 &  >31.24 &        &        &        &        &        \\
 62.7 &  84 &   29.44 &  55.41 &  83.06 &  34.59 &  50.94 & 167.51 \\
 63.1 &  86 &   18.53 &  55.05 &  82.82 &  34.95 &  51.18 & 167.60 \\
 63.6 &  88 &   12.78 &  54.34 &  82.37 &  35.66 &  51.63 & 170.29 \\
 64.0 &  90 &   13.13 &  55.82 &  84.31 &  34.18 &  49.69 & 163.16 \\
 64.4 &  92 &   13.06 &  57.16 &  86.88 &  32.84 &  47.12 & 153.31 \\
 64.9 &  94 &    2.79 &  68.98 & 104.89 &  21.02 &  29.11 & 140.16 \\
\colrule
 &  40-60 &  >30.97 &        &        &        &        &        \\
103.2 &  62 &   19.97 &  56.22 &  85.33 &  33.78 &  48.67 & 152.47 \\
103.6 &  64 &   12.61 &  57.83 &  87.12 &  32.17 &  46.88 & 149.35 \\
 104.1 &  66 &   10.02 &  58.08 &  87.99 &  31.92 &  46.01 & 145.88 \\
 104.5 &  68 &    4.47 &  66.76 & 101.33 &  23.24 &  32.67 & 128.07 \\
\colrule
 &  62-64 &  >31.17 &        &        &        &        &        \\
148.8 &  66 &    2.38 &  69.34 & 104.57 &  20.66 &  29.43 & 139.46 \\
  \end{tabular}
 \end{ruledtabular}
 \end{table}

 \begin{table}[ht]
 \caption{$^{50}\mbox{Ca}+^{176}\mbox{Yb}$ at $E_{c.m.}=172$~MeV. \label{tab:50Ca}}
 \begin{ruledtabular}
 \begin{tabular}{lccccccc}
$\beta$ (deg.) & $L$ ($\hbar$) & $\tau$ (zs)& $Z_H$ & $N_H$ & $Z_L$ & $N_L$ & TKE (MeV)\\
\colrule
 &  70-76 &  \textgreater31.39 &        &        &        &        &        \\
 16.4 &  78 &   28.37 &  53.48 &  81.82 &  36.52 &  54.18 & 164.50 \\
 16.8 &  80 &  \textgreater31.39 &        &        &        &        &        \\
 17.3 &  82 &	12.07 &  58.52 &  90.46 &  31.48 &  45.54 & 146.02 \\
 17.7 &  84 &	10.68 &  59.91 &  92.43 &  30.09 &  43.57 & 129.50 \\
 18.1 &  86 &	11.11 &  59.86 &  92.45 &  30.14 &  43.55 & 131.91 \\
 18.6 &  88 &	 5.10 &  66.46 & 102.23 &  23.54 &  33.77 & 132.02 \\
\colrule
 &  70-72 &  \textgreater31.22 &        &        &        &        &        \\
 60.3 &  74 &   33.88 &  54.50 &  83.29 &  35.50 &  52.71 & 163.56 \\
 60.7 &  76 &	26.42 &  54.83 &  83.80 &  35.17 &  52.20 & 163.95 \\
 61.1 &  78 &  >31.23 &        &	&	 &	  &	   \\
 61.6 &  80 &  >31.23 &        &	&	 &	  &	   \\
 62.0 &  82 &	22.91 &  54.93 &  83.87 &  35.07 &  52.13 & 166.80 \\
 62.4 &  84 &	16.19 &  53.94 &  82.41 &  36.06 &  53.59 & 167.72 \\
 62.8 &  86 &	17.64 &  53.37 &  82.37 &  36.63 &  53.63 & 167.93 \\
 63.3 &  88 &	17.73 &  54.75 &  83.06 &  35.25 &  52.94 & 167.16 \\
 63.7 &  90 &	12.21 &  54.45 &  83.31 &  35.55 &  52.69 & 165.07 \\
 64.1 &  92 &	11.29 &  57.94 &  88.92 &  32.06 &  47.08 & 148.43 \\
 64.6 &  94 &	 9.99 &  59.37 &  91.49 &  30.63 &  44.51 & 139.51 \\
 65.0 &  96 &	 3.00 &  68.59 & 105.93 &  21.41 &  30.07 & 137.42 \\
\colrule
 &  50-60 &  \textgreater30.96 &        &        &        &        &        \\
 103.0 &  62 &   13.33 &  56.90 &  87.23 &  33.10 &  48.77 & 148.65 \\
 103.4 &  64 &   14.79 &  57.56 &  88.64 &  32.44 &  47.36 & 143.92 \\
 103.8 &  66 &   12.85 &  56.58 &  86.97 &  33.42 &  49.03 & 143.38 \\
 104.3 &  68 &   12.36 &  58.21 &  89.65 &  31.79 &  46.35 & 145.12 \\
 104.7 &  70 &    4.90 &  65.88 & 100.66 &  24.12 &  35.34 & 132.51 \\
\colrule
 &  60-64 &  \textgreater31.17 &        &        &        &        &        \\
148.6 &  66 &   19.72 &  55.45 &  85.30 &  34.55 &  50.70 & 155.28 \\
149.0 &  68 &    2.81 &  68.55 & 105.29 &  21.45 &  30.71 & 135.83 
 \end{tabular}
 \end{ruledtabular}
 \end{table}

 \begin{table}[ht]
 \caption{$^{52}\mbox{Ca}+^{176}\mbox{Yb}$ at $E_{c.m.}=172$~MeV. \label{tab:52Ca}}
 \begin{ruledtabular}
 \begin{tabular}{lccccccc}
$\beta$ (deg.) & $L$ ($\hbar$) & $\tau$ (zs)& $Z_H$ & $N_H$ & $Z_L$ & $N_L$ & TKE (MeV)\\
\colrule
 &  50-66 &  >31.39 &        &        &        &        &        \\
 14.0 &  68 &   27.77 &  54.22 &  84.26 &  35.78 &  53.74 & 160.58 \\
 14.5 &  70 &  >31.39 &        &        &        &        &        \\
 14.9 &  72 &  >31.39 &        &        &        &        &        \\
 15.3 &  74 &   28.33 &  53.94 &  83.70 &  36.06 &  54.30 & 162.27 \\
 15.7 &  76 &   21.38 &  53.15 &  82.81 &  36.85 &  55.19 & 165.65 \\
 16.1 &  78 &   22.07 &  53.87 &  83.69 &  36.13 &  54.31 & 160.77 \\
 16.6 &  80 &   17.03 &  53.64 &  83.83 &  36.36 &  54.17 & 166.26 \\
 17.0 &  82 &   12.54 &  56.52 &  87.50 &  33.48 &  50.50 & 145.86 \\
 17.4 &  84 &   11.27 &  59.58 &  92.87 &  30.42 &  45.13 & 135.65 \\
 17.8 &  86 &	10.54 &  60.01 &  93.97 &  29.99 &  44.03 & 136.84 \\
 18.3 &  88 &	 5.95 &  64.70 & 100.76 &  25.30 &  37.24 & 132.61 \\
 18.7 &  90 &	 4.05 &  66.76 & 103.60 &  23.24 &  34.40 & 135.01 \\
\colrule
  &  50-62 &  >37.93 &        &        &        &        &        \\
 58.0 &  64 &   36.57 &  54.22 &  84.09 &  35.78 &  53.91 & 162.19 \\
 58.4 &  66 &  >37.95 &        &        &        &        &        \\
 58.8 &  68 &   28.94 &  54.09 &  83.98 &  35.91 &  54.02 & 164.16 \\
 59.2 &  70 &   27.22 &  54.08 &  83.78 &  35.92 &  54.22 & 162.65 \\
 59.6 &  72 &  >37.95 &        &        &        &        &        \\
 60.1 &  74 &   24.70 &  54.02 &  83.99 &  35.98 &  54.01 & 166.43 \\
 60.5 &  76 &   32.51 &  54.29 &  84.17 &  35.71 &  53.83 & 162.68 \\
 60.9 &  78 &   25.11 &  54.73 &  84.73 &  35.27 &  53.27 & 161.56 \\
 61.3 &  80 &   25.08 &  54.24 &  83.81 &  35.76 &  54.19 & 162.44 \\
 61.7 &  82 &   20.82 &  53.60 &  83.33 &  36.40 &  54.67 & 162.79 \\
 62.1 &  84 &   17.05 &  55.11 &  85.56 &  34.89 &  52.44 & 162.06 \\
 62.6 &  86 &   14.06 &  54.51 &  84.83 &  35.49 &  53.17 & 161.38 \\
 63.0 &  88 &   15.39 &  54.64 &  85.28 &  35.36 &  52.72 & 161.43 \\
 63.4 &  90 &   12.67 &  54.32 &  84.58 &  35.68 &  53.42 & 158.46 \\
 63.8 &  92 &   13.89 &  57.79 &  89.66 &  32.21 &  48.34 & 147.10 \\
 64.2 &  94 &   10.09 &  59.23 &  92.17 &  30.77 &  45.83 & 138.84 \\
 64.7 &  96 &    4.63 &  66.31 & 103.51 &  23.69 &  34.49 & 135.69 \\
 65.1 &  98 &    2.82 &  68.58 & 107.49 &  21.42 &  30.51 & 136.59 \\
\colrule
 &  40-58 &  >30.95 &        &        &        &        &        \\
102.4 &  60 &   16.29 &  54.62 &  85.04 &  35.38 &  52.96 & 157.27 \\
 102.8 &  62 &   14.37 &  56.33 &  87.52 &  33.67 &  50.48 & 148.15 \\
 103.2 &  64 &   12.84 &  55.90 &  87.05 &  34.10 &  50.95 & 149.30 \\
 103.6 &  66 &   19.34 &  57.33 &  89.50 &  32.67 &  48.50 & 141.76 \\
 104.0 &  68 &   15.91 &  57.78 &  89.52 &  32.22 &  48.48 & 138.51 \\
 104.5 &  70 &   12.09 &  56.42 &  88.03 &  33.58 &  49.97 & 140.88 \\
 104.9 &  72 &    4.68 &  66.05 & 102.59 &  23.95 &  35.41 & 128.11 \\
\colrule
 &  64-66 &  >44.67 &        &        &        &        &        \\
148.8 &  68 &   19.74 &  56.35 &  88.32 &  33.65 &  49.68 & 147.03 \\
149.2 &  70 &    3.02 &  68.29 & 106.45 &  21.71 &  31.55 & 132.02 \\
  \end{tabular}
 \end{ruledtabular}
 \end{table}

 \begin{table}[ht]
 \caption{$^{54}\mbox{Ca}+^{176}\mbox{Yb}$ at $E_{c.m.}=172$~MeV. \label{tab:52Ca}}
 \begin{ruledtabular}
 \begin{tabular}{lccccccc}
$\beta$ (deg.) & $L$ ($\hbar$) & $\tau$ (zs)& $Z_H$ & $N_H$ & $Z_L$ & $N_L$ & TKE (MeV)\\
\colrule
 &  50-60 &  >33.64 &        &        &        &        &        \\
 12.6 &  62 &   28.03 &  54.33 &  85.14 &  35.67 &  54.86 & 161.34 \\
 13.0 &  64 &  >33.64 &        &        &        &        &        \\
 13.4 &  66 &  >33.64 &        &        &        &        &        \\
 13.8 &  68 &  >33.64 &        &        &        &        &        \\
 14.3 &  70 &  >33.64 &        &        &        &        &        \\
 14.7 &  72 &  >33.64 &        &        &        &        &        \\
 15.1 &  74 &   32.16 &  53.82 &  84.74 &  36.18 &  55.26 & 161.42 \\
 15.5 &  76 &   29.59 &  53.54 &  84.65 &  36.46 &  55.35 & 163.86 \\
 15.9 &  78 &   28.91 &  53.54 &  83.87 &  36.46 &  56.13 & 164.13 \\
 16.3 &  80 &   26.90 &  52.96 &  83.28 &  37.04 &  56.72 & 165.23 \\
 16.7 &  82 &   20.85 &  54.43 &  85.44 &  35.57 &  54.56 & 158.90 \\
 17.2 &  84 &   11.57 &  55.41 &  87.26 &  34.59 &  52.74 & 147.58 \\
 17.6 &  86 &    8.67 &  60.66 &  95.62 &  29.34 &  44.38 & 132.41 \\
 18.0 &  88 &   11.56 &  59.37 &  93.84 &  30.63 &  46.16 & 132.68 \\
 18.4 &  90 &   10.43 &  60.21 &  95.39 &  29.79 &  44.61 & 131.24 \\
 18.8 &  92 &    4.68 &  65.99 & 103.66 &  24.01 &  36.34 & 131.07 \\
\colrule
57.0 &   60 &   37.74 &  54.21 &  85.83 &  35.79 &  54.17 & 158.98 \\
 57.4 &  62 &  >46.92 &        &        &        &        &        \\
 57.8 &  64 &  >46.92 &        &        &        &        &        \\
 58.2 &  66 &   46.44 &  53.29 &  84.01 &  36.71 &  55.99 & 161.12 \\
 58.6 &  68 &   28.42 &  54.91 &  86.26 &  35.09 &  53.74 & 157.12 \\
 59.0 &  70 &   26.14 &  54.12 &  85.28 &  35.88 &  54.72 & 158.28 \\
 59.4 &  72 &   42.58 &  53.63 &  84.67 &  36.37 &  55.33 & 160.71 \\
 59.8 &  74 &   27.78 &  53.86 &  84.96 &  36.14 &  55.04 & 161.45 \\
 60.2 &  76 &   26.52 &  54.28 &  85.27 &  35.72 &  54.73 & 160.52 \\
 60.7 &  78 &   24.14 &  53.83 &  84.38 &  36.17 &  55.62 & 158.28 \\
 61.1 &  80 &   33.45 &  53.72 &  84.98 &  36.28 &  55.02 & 160.28 \\
 61.5 &  82 &   23.80 &  53.93 &  84.74 &  36.07 &  55.26 & 161.76 \\
 61.9 &  84 &   22.42 &  51.63 &  80.98 &  38.37 &  59.02 & 171.96 \\
 62.3 &  86 &   13.00 &  55.85 &  88.24 &  34.15 &  51.76 & 154.09 \\
 62.7 &  88 &   11.70 &  54.68 &  85.90 &  35.32 &  54.10 & 159.40 \\
 63.1 &  90 &   12.26 &  56.68 &  89.00 &  33.32 &  51.00 & 153.20 \\
 63.5 &  92 &   13.48 &  56.55 &  89.11 &  33.45 &  50.89 & 151.40 \\
 64.0 &  94 &   12.64 &  57.08 &  90.23 &  32.92 &  49.77 & 145.84 \\
 64.4 &  96 &    8.78 &  59.25 &  92.90 &  30.75 &  47.10 & 140.49 \\
 64.8 &  98 &    4.84 &  66.22 & 104.55 &  23.78 &  35.45 & 133.27 \\
 65.2 & 100 &    3.05 &  67.58 & 107.32 &  22.42 &  32.68 & 137.09 \\
\colrule
 &  42-58 &  >30.95 &        &        &        &        &        \\
102.2 &  60 &   23.64 &  54.55 &  86.07 &  35.45 &  53.93 & 156.26 \\
102.6 &  62 &   16.54 &  57.17 &  90.15 &  32.83 &  49.85 & 144.84 \\
103.0 &  64 &   17.31 &  57.82 &  91.21 &  32.18 &  48.79 & 144.28 \\
103.4 &  66 &   15.95 &  56.59 &  89.40 &  33.41 &  50.60 & 144.54 \\
103.8 &  68 &   12.64 &  56.59 &  89.24 &  33.41 &  50.76 & 145.83 \\
 104.3 &  70 &   14.55 &  57.34 &  90.32 &  32.66 &  49.68 & 137.54 \\
 104.7 &  72 &   14.35 &  58.36 &  92.17 &  31.64 &  47.83 & 137.37 \\
 105.1 &  74 &    7.79 &  63.86 & 100.22 &  26.14 &  39.78 & 130.38 \\
 105.5 &  76 &    3.18 &  67.31 & 106.01 &  22.69 &  33.99 & 128.87 \\
 \colrule
 &  66-68 &  >33.42 &        &        &        &        &        \\
149.0 &  70 &   16.44 &  56.25 &  89.30 &  33.75 &  50.70 & 143.38 \\
149.4 &  72 &    5.17 &  65.86 & 104.06 &  24.14 &  35.94 & 131.27 \\
149.8 &  74 &    2.21 &  69.07 & 108.59 &  20.93 &  31.41 & 139.28 
  \end{tabular}
 \end{ruledtabular}
 \end{table}

 \begin{table}[ht]
 \caption{$^{56}\mbox{Ca}+^{176}\mbox{Yb}$ at $E_{c.m.}=172$~MeV. \label{tab:56Ca}}
 \begin{ruledtabular}
 \begin{tabular}{lccccccc}
$\beta$ (deg.) & $L$ ($\hbar$) & $\tau$ (zs)& $Z_H$ & $N_H$ & $Z_L$ & $N_L$ & TKE (MeV)\\
\colrule
 &  50-58 &  >40.39 &        &        &        &        &        \\
 12.0 &  60 &   30.05 &  53.93 &  86.17 &  36.07 &  55.83 & 158.19 \\
 12.4 &  62 &  >40.38 &        &        &        &        &        \\
 12.8 &  64 &  >40.38 &        &        &        &        &        \\
 13.2 &  66 &  >40.38 &        &        &        &        &        \\
 13.6 &  68 &  >40.38 &        &        &        &        &        \\
 14.1 &  70 &  >40.38 &        &        &        &        &        \\
 14.5 &  72 &  >40.38 &        &        &        &        &        \\
 14.9 &  74 &   38.20 &  52.93 &  84.63 &  37.07 &  57.37 & 163.86 \\
 15.3 &  76 &   20.67 &  54.28 &  86.51 &  35.72 &  55.49 & 159.23 \\
 15.7 &  78 &   31.92 &  53.30 &  85.05 &  36.70 &  56.95 & 161.33 \\
 16.1 &  80 &   37.51 &  53.45 &  85.18 &  36.55 &  56.82 & 160.50 \\
 16.5 &  82 &   28.44 &  53.08 &  84.59 &  36.92 &  57.41 & 161.35 \\
 16.9 &  84 &	21.78 &  53.87 &  85.92 &  36.13 &  56.08 & 159.12 \\
 17.3 &  86 &	16.31 &  55.94 &  89.19 &  34.06 &  52.81 & 146.84 \\
 17.7 &  88 &	 8.84 &  59.08 &  94.27 &  30.92 &  47.73 & 140.13 \\
 18.2 &  90 &	 7.93 &  59.93 &  95.77 &  30.07 &  46.23 & 136.00 \\
 18.6 &  92 &	 8.52 &  61.10 &  98.17 &  28.90 &  43.83 & 129.96 \\
 19.0 &  94 &	 8.37 &  60.99 &  98.02 &  29.01 &  43.98 & 134.29 \\
 19.4 &  96 &	 4.93 &  65.43 & 104.12 &  24.57 &  37.88 & 131.71 \\
\colrule
 &  50-64 &  >35.67 &        &        &        &        &        \\
 58.0 &  66 &   31.12 &  52.98 &  84.65 &  37.02 &  57.35 & 161.08 \\
 58.4 &  68 &  >35.66 &        &        &        &        &        \\
 58.8 &  70 &   32.22 &  53.92 &  85.58 &  36.08 &  56.42 & 157.15 \\
 59.2 &  72 &   28.58 &  54.04 &  86.09 &  35.96 &  55.91 & 157.06 \\
 59.6 &  74 &   28.68 &  52.72 &  84.02 &  37.28 &  57.98 & 166.58 \\
 60.0 &  76 &   19.71 &  54.06 &  86.40 &  35.94 &  55.60 & 161.40 \\
 60.4 &  78 &   16.95 &  54.49 &  87.18 &  35.51 &  54.82 & 161.48 \\
 60.8 &  80 &   20.74 &  53.41 &  85.01 &  36.59 &  56.99 & 159.48 \\
 61.2 &  82 &   25.67 &  53.21 &  84.30 &  36.79 &  57.70 & 158.36 \\
 61.6 &  84 &   22.25 &  52.39 &  83.43 &  37.61 &  58.57 & 163.97 \\
 62.1 &  86 &   21.28 &  53.43 &  84.50 &  36.57 &  57.50 & 159.69 \\
 62.5 &  88 &   12.11 &  54.26 &  86.35 &  35.74 &  55.65 & 158.45 \\
 62.9 &  90 &   19.64 &  55.04 &  87.08 &  34.96 &  54.92 & 159.07 \\
 63.3 &  92 &   11.50 &  54.72 &  87.75 &  35.28 &  54.25 & 158.68 \\
 63.7 &  94 &   13.02 &  54.48 &  86.58 &  35.52 &  55.42 & 151.00 \\
 64.1 &  96 &   11.16 &  56.41 &  90.26 &  33.59 &  51.74 & 150.04 \\
 64.5 &  98 &   10.18 &  57.09 &  91.49 &  32.91 &  50.51 & 136.60 \\
 64.9 & 100 &    7.38 &  60.71 &  97.18 &  29.29 &  44.82 & 141.40 \\
 65.3 & 102 &    5.34 &  65.33 & 104.33 &  24.67 &  37.67 & 130.42 \\
 65.8 & 104 &    3.11 &  67.09 & 107.79 &  22.91 &  34.21 & 138.68 \\
\colrule
 &  44-60 &  >33.19 &        &        &        &        &        \\
102.4 &  62 &   27.70 &  53.21 &  85.09 &  36.79 &  56.91 & 158.34 \\
102.8 &  64 &   18.72 &  56.46 &  90.52 &  33.54 &  51.48 & 145.72 \\
103.2 &  66 &   12.90 &  56.16 &  89.81 &  33.84 &  52.19 & 146.26 \\
103.6 &  68 &   11.21 &  56.16 &  89.63 &  33.84 &  52.37 & 143.27 \\
104.1 &  70 &   12.89 &  56.45 &  89.80 &  33.55 &  52.20 & 147.59 \\
104.5 &  72 &	10.03 &  57.40 &  91.61 &  32.60 &  50.39 & 143.40 \\
104.9 &  74 &	14.24 &  57.49 &  92.09 &  32.51 &  49.91 & 134.80 \\
105.3 &  76 &	10.38 &  59.84 &  95.34 &  30.16 &  46.66 & 132.34 \\
105.7 &  78 &	 5.58 &  64.64 & 103.23 &  25.36 &  38.77 & 128.89 \\
106.1 &  80 &	 3.07 &  66.34 & 105.60 &  23.66 &  36.40 & 141.78 \\
\colrule
 &  68-70 &  >31.16 &        &        &        &        &        \\
149.2 &  72 &   20.09 &  53.54 &  85.14 &  36.46 &  56.86 & 162.40 \\
149.6 &  74 &   12.97 &  57.70 &  92.07 &  32.30 &  49.93 & 137.27 \\
150.0 &  76 &    3.10 &  67.22 & 106.91 &  22.78 &  35.09 & 136.21 \\
 \end{tabular}
 \end{ruledtabular}
 \end{table}

\clearpage


\end{document}